\documentclass{revtex4}
\def\[{\left\lbrack}
\def\]{\right\rbrack}

\def\({\left(}
\def\){\right)}
\newcommand{\be}{\begin{equation}}
\newcommand{\ee}{\end{equation}}
\newcommand{\ea}{\end{eqnarray}}
\newcommand{\ba}{\begin{eqnarray}}

\begin{document}

\title{The Dual of the Carroll-Field-Jackiw Model}

\author{M. S. Guimar\~aes, L. Grigorio, C. Wotzasek}
\email{marcelosg, grigorio, clovis@if.ufrj.br}
\affiliation{Instituto de F\'{\i}sica, Universidade Federal do Rio de Janeiro,\\
21945-970, Rio de Janeiro, RJ, Brasil}

\begin{abstract}
\vspace{0.4in}
\begin{center}
{\bf Abstract}
\end{center}
\noindent In this work we apply different duality techniques, both the dual
projection, based on the soldering formalism and the master action, in order to
obtain and study the dual description of the Carroll-Field-Jackiw model \cite{cfj}, a
theory with a Chern-Simons-like explicitly Lorentz and CPT violating term, including
the interaction with external charges. This Maxwell-Chern-Simons-like model may be
rewritten in terms of the interacting modes of a massless scalar model and a
topologically massive model \cite{mcs}, that are mapped, through duality, into
interacting massless Maxwell and massive self-dual modes \cite{sd}. It is also shown
that these dual modes might be represented into an unified rank-two self-dual model
that represents the direct dual of the vector Maxwell-Chern-Simons-like model.

\end{abstract}

\maketitle


\setlength{\baselineskip} {20 pt}

\vspace{0.7in}
\section{Introduction}
Quantum field theories that describe the physical reality as we know it are all
Lorentz and CPT invariant and the validity of these symmetries have been confirmed by
various experiments. Nevertheless the search for a quantum description of gravity has
led us to believe that fundamental changes in the structure of space-time occur at
the Planck scale \cite{solvay}. A violation of Lorentz symmetry is expected and in
fact in string theory there is the possibility of the condensation (i.e. development
of a vacuum expectation value) of tensor fields leading to a spontaneous breaking of
the Lorentz invariance \cite{ks}.

Recently many works \cite{testsLV} have treated the possible effects that a violation
of the Lorentz symmetry has at low-energy, possibly presently observed, scales. The
seminal work of Colladay and Kostelecky \cite{colkostESM} addressed the possible
modifications to the Standard Model, the resultant model being named Extended Standard
Model (ESM). Of particular interest to us in this work is the QED sector of the ESM,
more precisely, the CPT-odd pure-photon sector.

Carroll, Field and Jackiw have tackled this problem in \cite{cfj} by studying the
action
\begin{eqnarray}
\label{CFJ}
  {\cal L}_{CFJ} &=& - \frac 14 F_{\mu\nu}F^{\mu\nu} +
  p_{\mu}\varepsilon^{\mu\nu\rho\sigma}A_{\nu}\partial_{\rho}A_{\sigma}\, ,
\end{eqnarray}
where $F_{\mu\nu} = \partial_{\mu}A_{\nu}-\partial_{\nu}A_{\mu}$ and we work with the
metric signature $(+,-,-,-)$. The Chern-Simons like term added to the usual Maxwell
Lagrangian is responsible for the breaking of Lorentz and CPT invariance. Lorentz
symmetry is broken because $p_{\mu}$, being a constant four-vector, selects a
preferred direction in space-time for each Lorentz frame.

Some authors have explored the physical aspects of this model \cite{soldati, klink}.
As a theory for a modified electromagnetism it has been shown that the vacuum becomes a
birefringent media, and it was realized that this effect could be used to set limits
in the magnitude of the Lorentz violating vector $p_{\mu}$. In \cite{cfj} only a
time-like $p_{\mu}$ was considered and it was argued that astrophysical observations
of polarized light and geomagnetic data seems to rule out a non-vanishing magnitude
of $p_{\mu}$ in this case. For the space-like case, astronomical observations
\cite{nodland} were used to argue in favor of a non-vanishing value of the magnitude
of $p_{\mu}$ but the results has been disputed \cite{cohen}.

Discussions concerning the consistence of the quantum field theory (QFT) defined by
(\ref{CFJ}) as a function of the Lorentz character of $p_{\mu}$ has also been carried
out \cite{soldati, klink}. It was noted that a time-like $p_{\mu}$ gives rise to a
QFT for which unitarity and microcausality cannot be satisfied simultaneously. On the
other hand it seems that a consistent QFT can be defined for a space-like $p_{\mu}$.
These features are already apparent in the dispersion relation as we shall briefly
recall in section 3.

The purpose of the present study is to work out the dual version of the CFJ model in order to gain some
further insights about the physical nature of the dynamical modes of the theory (\ref{CFJ}) and its symmetries. To this end we shall employ the dual projection technique that has been used before to separate out the dynamical modes from the symmetry sector of a given model
\footnote{This technique has been developed in the context of the soldering formalism, to study the physical contents of many different models in distinct dimensions \cite{Abreu:1998tc,Abreu:2004xe,Wotzasek:2000wy,Banerjee:2000fx,Wotzasek:1998rj}.}.
In $(2+1)D$ a Chern-Simons term can be naturally added
to the Maxwell theory (without the vector $p_{\mu}$ and hence maintaining Lorentz
invariance, since the Levi-Civita symbol $\varepsilon$ has $3$ indices), the resultant theory, known as the Maxwell-Chern-Simons (MCS) or topologically massive theory, was extensively studied \cite{mcs} and is related with planar electromagnetic phenomena such as the quantum Hall effect. It was
realized that this theory has a dual description \cite{dj} in terms of another local
vector field theory without gauge symmetry known as the self-dual (SD) model
\cite{sd} (we will briefly review this result in the following section). The question
we pose is if this kind of dual relation exists for the CFJ model. We will show that
a dual description indeed exists and we hope that the result might shed some light on
the properties of this model.

The paper is organized as follows. In the next section we will review the MCS-SD
duality relation and present the techniques that will be used to discuss the CFJ
model. In section 3 we will discuss the CFJ model recalling some of its physical
properties and find its dual formulation applying the methods reviewed in section 2.
In section 4 we extend our results for the case in which the model is minimally
coupled to a conserved external current. This will be done by writing the current in
terms of a so called Chern-kernel. It will be interesting to see how the Chern-kernel
naturally leads an electric coupling into a magnetic coupling as we go through the duality procedure. Finally in section 5 we present our conclusions.

\vspace{0.4in}

\section{Remembering The Maxwell-Chern-Simons/Self-dual Duality}

Before engaging into the discussion of the duality of the CFJ-model, it is interesting to review some general known facts about duality in electromagnetic-like theories, such as massless $p$-form Maxwell   theories, massive Proca-like and topologically massive theories. For the first two, the duality transformation and the structure of the corresponding duality groups have long been the focus of detailed studies\cite{Deser:1997mz,Wotzasek:1998rj,Noronha:2003vp}. In Maxwell-like theories, the ranks of the dual pair $A_p$ and $\mbox{}^*{A_q}$ obey the relation $p+q+2=D$, coming from the restrictions imposed by the Gauss constraints. The massive duality for the Proca-like fields obey a similar relation as $p+q+1=D$. A similar systematic study for topologically massive and self-dual models for $p$-form potentials in arbitrary dimensions has appeared more recently showing that these fields obey the latter rule albeit the presence of the gauge symmetry\cite{Menezes:2003vz}. In this section we shall review the the basic concepts for the $D=3$ topologically massive model using the dual projection technique and the standard dualization based on the master action.

The dual technique, known as dual projection, is quite useful to separate the dynamical content of a field theory from its symmetry carrying part. In this way it may help to shed light over the physical significance of a model where other approachs fail. For instance, in $D=2$ this technique was used to study the different formulations of the chiral boson models \cite{Siegel:1983es,Floreanini:1987as} and establish their equivalence\cite{Abreu:1998tc}. We found that the chiral dynamics of the Siegel's model is carryied by Floreanini-Jackiw fields while the chiral diffeomorphism symmetry is represented by a Hull noton\cite{Abreu:2002gc}. More recently, the situation with Lorentz symmetry breaking was also considered under this approach\cite{Abreu:2004xe}. In $4D$ Maxwell-like theories this analysis was considered in \cite{Wotzasek:1998rj} mostly to display the structure of the duality group and it was extended to the massive case in \cite{Noronha:2003vp}. In $3D$ models, the dynamics of topologically massive theory where analysed through the contents of their dual versions that made explicit their dynamical and symmetry contents through a dual projection of their original fields\cite{Wotzasek:2000wy}
\cite{Banerjee:2000fx}. Other applications may be found in \cite{Dalmazi:2003sz,Dalmazi:2005bp,Dalmazi:2006yv,Banerjee:2000hs,Banerjee:2000mr}.

In this section we will recall some results concerning the dual equivalence between
Maxwell-Chern-Simons (MCS) theory \cite{mcs}
\begin{eqnarray}
  {\cal L}_{MCS}= -\frac 14 F_{\mu\nu}F^{\mu\nu} + \frac m2 A_{\mu}\varepsilon^{\mu\nu\rho}\partial_{\nu}A_{\rho},
  \label{mcsad01}
\end{eqnarray}
where $F_{\mu\nu} = \partial_{\mu}A_{\nu} - \partial_{\nu}A_{\mu}$, and the self-dual
(SD) model \cite{sd}
\begin{eqnarray}
  {\cal L}_{SD}= \frac 12 f_{\mu}f^{\mu} - \frac {1}{2m} f_{\mu}\varepsilon^{\mu\nu\rho}\partial_{\nu}f_{\rho}.
  \label{mcsad02}
\end{eqnarray}
Both models describe a topologically massive excitation of spin $1$ in $(2+1)D$. The
interesting point is that the MCS theory has a gauge symmetry ($A_{\mu} \rightarrow
A_{\mu} + \partial_{\mu}\Lambda$) but in the SD theory this symmetry is absent.
Nevertheless it was shown that those two models describe the same propagating degrees
of freedom \cite{dj}.

The basic idea in the duality transformation known as dual projection is to perform a canonical transformation leading the separation of the explicit degrees of freedom, when possible.
In this case, consider the lagrangian density (\ref{mcsad01}), in its first-order form,
\begin{eqnarray}
 {\cal L}_{M} &=& \Pi_{\mu}(\varepsilon^{\mu\nu\rho}\partial_{\nu}A_{\rho}) + \frac 12 \Pi_{\mu}\Pi^{\mu} + \frac m2 A_{\mu}\varepsilon^{\mu\nu\rho}\partial_{\nu}A_{\rho}\nonumber\\
              &=&(\frac m2 A_{\mu} + \Pi_{\mu})(\varepsilon^{\mu\nu\rho}\partial_{\nu}A_{\rho}) + \frac 12 \Pi_{\mu}\Pi^{\mu},
    \label{mcsadmestra}
\end{eqnarray}
where $\Pi_\mu$ is an auxiliary vector-field which may be integrated out to give us back the MCS
lagrangian. By making the redefinition $\frac m2 A_{\mu} + \Pi_{\mu} = B_{\mu}$, we
find
\begin{eqnarray}
{\cal L}_{M} = B_{\mu}(\varepsilon^{\mu\nu\rho}\partial_{\nu}A_{\rho}) + \frac 12
(B_{\mu} - \frac m2 A_{\mu})(B^{\mu} - \frac m2 A^{\mu}).
    \label{mcsad011}
\end{eqnarray}
Observe that by definition $B_{\mu}$ transforms as $B_{\mu} \rightarrow B_{\mu} +
\frac m2 \partial_{\mu}\Lambda$ whenever $A_{\mu} \rightarrow A_{\mu} +
\partial_{\mu}\Lambda$, so that the gauge character of the $A_\mu$ field has not changed.
Next we may perform a canonical transformation in the space of the fields to reveal the self-dual and pure gauge nature of the components
\begin{eqnarray}
 B_{\mu} &=& \frac 12 (A^{+}_{\mu} - A^{-}_{\mu})\nonumber\\
   A_{\mu} &=& \frac 1m (A^{+}_{\mu} + A^{-}_{\mu}),
    \label{mcsad022}
\end{eqnarray}
which gives us
\begin{eqnarray}
{\cal L}_{M} = \frac{1}{2m}
A^{+}_{\mu}\varepsilon^{\mu\nu\rho}\partial_{\nu}A^{+}_{\rho} - \frac{1}{2m}
A^{-}_{\mu}\varepsilon^{\mu\nu\rho}\partial_{\nu}A^{-}_{\rho} + \frac 12
A^{-}_{\mu}A^{-\mu},
    \label{mcsad03}
\end{eqnarray}
or, renaming $A^{-}_{\mu} = f_{\mu}$ and $A^{+}_{\mu} = A_{\mu}$
\begin{eqnarray}
 {\cal L}_{M} = \frac{1}{2m} A_{\mu}\varepsilon^{\mu\nu\rho}\partial_{\nu}A_{\rho} +  \frac 12 f_{\mu}f^{\mu} - \frac{1}{2m} f_{\mu}\varepsilon^{\mu\nu\rho}\partial_{\nu}f_{\rho}  = {\cal L}_{CS} + {\cal L}_{SD},
    \label{mcsad04}
\end{eqnarray}
The first term is a topological term (the associated hamiltonian being null) and
the remaining terms we recognize as the SD model (\ref{mcsad02}).
It is clear by this procedure that the kind of gauge symmetry carried by the MCS
theory have completly innocuous dynamical character as is well know by the
properties of the pure Chern-Simons theory \cite{9902115}. As we were able to
separate this term we might say that the only information that the MCS theory has,
which is not present in the SD model, regards the topological character of the space
in which the theory is defined. It is very interesting and highly nontrivial that this
separation is possible. It means, for example, that the energy propagating modes is given
entirely by the SD sector in (\ref{mcsad04}) but the energy eigenstates have a
degeneracy parameterized by the Hilbert space of pure Chern-Simons theory (a
topological degeneracy) \cite{9304081}.

We can establish this duality by yet another technique that, although more direct, does not preserve the gauge structure of the original theory. Starting again with the
first-order lagrangian density (\ref{mcsadmestra}) we can eliminate the $A_{\mu}$ field using its
equation of motion
\begin{eqnarray}
 \varepsilon^{\mu\nu\rho}\partial_{\nu}\Pi_{\rho} =
 m\varepsilon^{\mu\nu\rho}\partial_{\nu}A_{\rho},
    \label{mcsad05}
\end{eqnarray}
which implies
\begin{eqnarray}
 \Pi_{\mu} = m A_{\mu} + \partial_{\mu}\phi,
    \label{mcsad06}
\end{eqnarray}
and the resulting lagrangian for $\Pi_{\mu}$ must comprise all the original dynamics
of $A_{\mu}$. However, by the very nature of this procedure (which seeks an alternative
description just for the \textbf{dynamics}) we loose information about the gauge
symmetry and consequently of the topological structure of the MCS theory. Indeed,
from (\ref{mcsad06}), $\Pi^{\mu}$ stands for a whole gauge orbit of $A^{\mu}$ and as
we substitute for $A^{\mu}$ in (\ref{mcsadmestra}) all the gauge freedom is lost while
the dynamical character is contained in the resulting self-dual theory
\begin{eqnarray}
 {\cal L}_{M} \rightarrow  {\cal L}_{SD} = \frac 12 f_{\mu}f^{\mu} - \frac{1}{2m} f_{\mu}\varepsilon^{\mu\nu\rho}\partial_{\nu}f_{\rho},
    \label{mcsad07}
\end{eqnarray}

This duality can be generalized to deal with the presence of couplings to external charges\cite{nosso,
anacleto} in great generality. We discuss the case of the MCS coupled minimally with a external source in
the appendix A. It represents a simpler example of the method we will use in section
IV to deal with the CFJ model.

\section{The dual of the Carroll-Field-Jackiw theory}

Now we can apply what we have learned in the previous section to the CFJ theory
defined by (\ref{CFJ}). However, before we engage in the search for the dual formulation
let us discuss some of the physical content of this theory that are already apparent
in its dispersion relation. Considering plane wave solutions with wave four-vector
$k^{\mu} = (\omega, \bf{k})$ we have \cite{cfj}
\begin{eqnarray}
 k^4 + k^2p^2 - (k\cdot p)^2 = 0.
    \label{dr}
\end{eqnarray}
Consider first a time-like Lorentz violating vector. As was already mentioned, this
situation is problematic\cite{soldati}. To see this we may consider for simplicity a frame in which
$p^{\mu} = (m,0,0,0)$ with $m>0$. The dispersion relation becomes
\begin{eqnarray}
 \omega^2 = {\bf k}^2 \pm m|{\bf k}|.
    \label{drtl}
\end{eqnarray}
The plus and minus sign correspond to two circularly polarized modes. The separation
of those two modes moving with different velocities is a clear sign of the Lorentz
violation. The minus sign mode has an imaginary energy at $|{\bf k}|<m$. In the
quantum theory, in order to preserve unitarity, this tachyonic mode has to be
excluded. But it has been shown in \cite{klink} that the exclusion of the region
$|{\bf k}|<m$ leads to a violation of microcausality. Because of these problems we
will focus our attention in the space-like $p^{\mu}$ from now on (some further comments on
the time-like $p^{\mu}$ case will be given below).

Considering a space-like $p^{\mu}$, we can work in a frame in which $p^{\mu} =
(0,0,0,m)$, $m>0$. The dispersion relation now reads
\begin{eqnarray}
 \omega^2 = {\bf k}^2 + \frac{m^2}{2} \pm \frac m2 \sqrt{4k^2_3 + m^2}\;\; ; \;\;
 {\bf k}^2 = k_1^2 + k_2^2 + k_3^2\, .
    \label{drsl}
\end{eqnarray}
These two degrees of freedom seem to have a consistent definition as excitations in
a quantum field theory as was discussed in \cite{klink} without problems of causality
or unitarity violations. In fact it can easily be seen that the velocity for
propagation of signals never exceeds the velocity of light.

As for the nature of those excitations we can readily see that if the dynamics is
restricted to happen in the plane perpendicular to the direction defined by the
space-like Lorentz-violating vector (in this case the $x^3$ direction), that is, if
we put $k_3 = 0$ in (\ref{drsl}), the plus sign corresponds to a massive excitation
of mass $m$ and the minus sign to a massless excitation. We can have a more precise
idea of these excitations by considering small variations out of the $x_1-x_2$ plane. If
$\frac {k_3}{m} \ll 1$, then
\begin{eqnarray}
\label{dr2}
  k^2 \approx \frac{m^2}{2} [ 1 \pm (1 + 2 \frac{k^{2}_3}{m^2})] =
  \left\{
              \begin{array}{ll}
                   m^2 + k^{2}_3, \;\;\mbox{massive case}\\
                   -k^{2}_3, \;\;\mbox{massless case}.
              \end{array}
       \right.
\end{eqnarray}
observe that for the massless case, to first order in $k_3$ the dispersion relation
is not modified
\begin{eqnarray}
\label{dr3}
  k^2 \approx -k^{2}_3 \Rightarrow \omega^2 = k^{2}_1 + k^{2}_2
\end{eqnarray}

What are the specific dynamics of these particles? This can be answered by
rearranging the CFJ Lagrangian (\ref{CFJ}). Singling out the $A_3$ component of the
$A_{\mu}$ field, renaming it $A^3 = \phi$ and introducing Latin indices $a,b,c = 0,1,2$
to denote the dynamics in the transverse plane, the lagrangian (\ref{CFJ}) becomes
\begin{eqnarray}
\label{CFJmod}
  {\cal L}_{CFJ} &=& - \frac 14 F_{ab}F^{ab} +
  m\varepsilon^{abc}A_{a}\partial_{b}A_{c} + \frac 12 \partial_a\phi\partial^a\phi +
  \phi\partial_a\partial_3A^a + \frac 12 \partial_3A_a\partial_3A^a\nonumber\\
  &=& {\cal L}_{MCS}(A) + {\cal L}_{scalar}(\phi) + \phi\partial_a\partial_3A^a + \frac 12
  \partial_3A_a\partial_3A^a.
\end{eqnarray}
If there is no $x^3$ dependence we indeed have two decoupled degrees of freedom lying
in the $x_1-x_2$ plane. The massless excitation is a scalar particle and the massive
excitation turns out to be a topologically massive particle described by a MCS
Lagrangian. It is known that a massless scalar particle in $(2+1)D$ has a dual
description in terms of Maxwell fields and, as we have seen, the MCS theory is dual to
the SD theory. We can tell immediately then that the dual description of
(\ref{CFJmod}), in this dimensionally reduced situation, should read
\begin{eqnarray}
\label{CFJ2Ddual}
  {\cal L}_{CFJ}\mid_{k_3 = 0} \rightarrow {\cal L}_{CS} + {\cal L}_{SD} + {\cal L}_{Maxwell}\, ,
\end{eqnarray}
where the pure Chern-Simons lagrangian does not describe a propagating degree of
freedom as we remarked in the previous section but carries part of the gauge symmetry
of the original theory.

Reintroducing the $x^3$ dependence, Eq.(\ref{drsl}) shows a distortion of the
energy-momentum relation and this is just a reflection of the last two terms in
(\ref{CFJmod}). Indeed this distortion is generated by the $k_3$ component of the
momentum, and those extra terms depend on $x^3$ derivatives.

Note also that in the reduced transverse plane, $A_a$ alone is a gauge field -- the potential for the MCS theory. In
$D=(3+1)$, from Eq.(\ref{CFJmod}) we see that only the combination $(A^a, \phi)$ (which is just the
original $A^{\mu}$) can be identified as a gauge field. It will be interesting to
note that in the dual version of (\ref{CFJmod}) the dual fields (which in the plane
are described by (\ref{CFJ2Ddual})) retain their identities as gauge or non-gauge
fields as we turn on or off the $x^3$ dependence. This dual description of
(\ref{CFJmod}) is what we seek now.

Consider the first-order Lagrangian obtained from (\ref{CFJ}) just by the introduction of
a rank-two auxiliary field ($\Pi_{\mu\nu}$)
\begin{eqnarray}
 {\cal L}_{MCFJ} = \frac 12 \Pi_{\mu\nu}\varepsilon^{\mu\nu\rho\sigma}\partial_{\rho}A_{\sigma} - \frac 14 \Pi_{\mu\nu}\Pi^{\mu\nu} +
 p_\mu\varepsilon^{\mu\nu\rho\sigma}A_{\nu}\partial_{\rho}A_{\sigma}.
    \label{mestraCFJ}
\end{eqnarray}
In what follows we will work with $p^{\mu} = (0,0,0,m)$, $m>0$, and again use Latin indices of the beginnig of the alphabet to denote coordinates in the transverse plane. Then, we can write (\ref{mestraCFJ}) as
\begin{eqnarray}
 {\cal L}_{MCFJ} = -\frac 12 A_{3}\varepsilon^{abc}\partial_{a}\Pi_{bc} + \frac 12 A_{a}\varepsilon^{abc}\partial_{3}\Pi_{bc} + A_{a}\varepsilon^{abc}\partial_{b}\Pi_{c3}- \frac 14 \Pi_{\mu\nu}\Pi^{\mu\nu} +
 m\varepsilon^{abc}A_{a}\partial_{b}A_{c}.
    \label{mestraCFJ2}
\end{eqnarray}
where a partial integration was performed and we have defined $\varepsilon^{abc3}
\equiv \varepsilon^{abc}$. Observe that $A_3$ is a Lagrange multiplier that enforces the constraint
\begin{eqnarray}
 \varepsilon^{abc}\partial_{a}\Pi_{bc} = 0,
    \label{lagmult}
\end{eqnarray}
whose solution, in terms of a new gauge potential $B_a$, is
\begin{eqnarray}
  \Pi_{ab}= \partial_{a}B_{b} -
 \partial_{b}B_{a} \equiv G_{ab}\, .
    \label{lagmult2}
\end{eqnarray}
We are then left with
\begin{eqnarray}
 {\cal L}_{MCFJ} &=& \frac 12 A_{a}\varepsilon^{abc}\partial_{3}G_{bc} - A_{a}\varepsilon^{abc}\partial_{b}h_{c}- \frac 14 G_{ab}G^{ab}
 + \frac 12 h_ah^a + m\varepsilon^{abc}A_{a}\partial_{b}A_{c}\nonumber\\
               &=& (-h_a + \partial_{3}B_a + mA_a)\varepsilon^{abc}\partial_{b}A_{c}- \frac 14 G_{ab}G^{ab}
 + \frac 12 h_ah^a.
    \label{mestraCFJ3}
\end{eqnarray}
where we have renamed the independent field $\Pi^{a3} \equiv h^a $. Following the steps of section 2 we define
\begin{eqnarray}
-h_a + mA_a = C_a
    \label{redef}
\end{eqnarray}
to obtain
\begin{eqnarray}
 {\cal L}_{MCFJ} \rightarrow C_a\varepsilon^{abc}\partial_{b}A_{c} + \partial_{3}B_a\varepsilon^{abc}\partial_{b}A_{c} - \frac 14 G_{ab}G^{ab}
 + \frac 12 (mA_a - C_a)(mA^a - C^a).
    \label{mestraCFJ4}
\end{eqnarray}
Observe that $C_a$ inherits the gauge symmetry of $A_a$ so that the last term above
is symmetric under $A_a \rightarrow A_a + \partial_a\phi$ with $C_a \rightarrow C_a +
m\partial_a\phi$. Now we perform a canonical transformation that reveals the dynamical contents and the symmetry structure of the model
\begin{eqnarray}
C_{a} &=& \frac 12 (A^{+}_{a} - A^{-}_{a})\nonumber\\
   A_{a} &=& \frac{1}{2m} (A^{+}_{a} + A^{-}_{a}),
    \label{rot}
\end{eqnarray}
which gives us
\begin{eqnarray}
 {\cal L}' = \frac{1}{4m} A^{+}_a\varepsilon^{abc}\partial_{b}A^{+}_{c} - \frac{1}{4m} A^{-}_a\varepsilon^{abc}\partial_{b}A^{-}_{c} + \frac{1}{2m}\partial_{3}B_a\varepsilon^{abc}\partial_{b}A^{+}_{c} + \frac{1}{2m}\partial_{3}B_a\varepsilon^{abc}\partial_{b}A^{-}_{c} - \frac 14 G_{ab}G^{ab}
 + \frac 12 A^{-}_a A^{-a}.
    \label{mestraCFJ5}
\end{eqnarray}
A further field redefinition is necessary to separated out the pure Chern-Simons term
\begin{eqnarray}
A^{+}_{a} + \partial_{3}B_a = D_a \, ,
    \label{redef2}
\end{eqnarray}
which leads us finally to (with $f_a = A^{-}_a$)
\begin{eqnarray}
 {\cal L}_{dualCFJ} = \frac{1}{4m} D_a\varepsilon^{abc}\partial_{b}D_{c} - \frac{1}{4m} f_a\varepsilon^{abc}\partial_{b}f_{c} + \frac 12 f_a f^a - \frac 14 G_{ab}G^{ab} + \frac{1}{2m} f_a\varepsilon^{abc}\partial_{b}\partial_{3}B_{c} -
 \frac{1}{4m}\partial_{3}B_a\varepsilon^{abc}\partial_{b}\partial_{3}B_{c}.
    \label{dualCFJ}
\end{eqnarray}
The first four terms are defined on the plane transverse to the Lorentz-breaking direction, the first being a pure Chern-Simons carrying part of the gauge symmetry while the second and third carry the self-dual dynamics. The fourth term is a pure Maxwell carrying the remaing gauge symmetry. The last two terms, as before in the CFJ-model, correspond to the interactions in the Lorentz-breaking direction.
This theory has the same physical content as the CFJ theory for a space-like
$p^{\mu}$ (to achieve orientations other than $p^{\mu} = (0,0,0,m)$ we just rotate
the coordinates). As anticipated, without $x^3$ dependence we obtain
(\ref{CFJ2Ddual}) with $D^a$ the pure Chern-Simons field, $f^a$ the SD field and
$B^a$ the Maxwell field. These retain their identities as gauge ($D^a$ and $B^a$) or
non-gauge ($f^a$) fields even in the full dimensional theory (\ref{dualCFJ}), a
feature not shared by the $\phi$ and $A^a$ fields in (\ref{CFJmod}) as was discussed
before.

We have done our calculations with the particular choice $p^{\mu} = (0,0,0,m)$ for
the sake of simplicity but it can be shown that the same procedures can be carried
out for a general, non-light-like ($p^2 \neq 0$), $p^{\mu}$. The result amounts to
the following expected generalizations (see appendix B for details)
\begin{eqnarray}
 \varepsilon^{abc}\partial_{b} &\rightarrow&  \varepsilon^{\mu\nu\rho\sigma}p_{\nu}\partial_{\rho}\nonumber\\
 \partial_3 &\rightarrow& p^{\mu}\partial_{\mu}.
    \label{corgenp}
\end{eqnarray}
So that the dual in this general case is given by
\begin{eqnarray}
 {\cal L}_{dualCFJ} &=& -\frac{1}{4p^4} D_{\mu}\varepsilon^{\mu\nu\rho\sigma}p_{\nu}\partial_{\rho}D_{\sigma} + \frac{1}{4p^4} f_{\mu}\varepsilon^{\mu\nu\rho\sigma}p_{\nu}\partial_{\rho}f_{\sigma} - \frac{1}{2p^2} f_{\mu} f^{\mu} - \frac{1}{2p^2} (\varepsilon^{\mu\nu\rho\sigma}p_{\nu}\partial_{\rho}B_{\sigma})^2\nonumber\\
  &+& \frac{1}{2p^4}
  f_{\mu}\varepsilon^{\mu\nu\rho\sigma}p_{\nu}\partial_{\rho}[(p^{\alpha}\partial_{\alpha})B_{\sigma}]
 + \frac{1}{4p^4}[(p^{\alpha}\partial_{\alpha})B_{\mu}]\varepsilon^{\mu\nu\rho\sigma}p_{\nu}\partial_{\rho}[(p^{\beta}\partial_{\beta})B_{\sigma}].
    \label{dualCFJgen}
\end{eqnarray}
Observe that the components of the fields along the direction defined by $p^{\mu}$
are effectively null, this is the generalization of the fact that in (\ref{dualCFJ})
there is no $3$-component in any field.

The separation of the original dynamical contents of the field $A_{\mu}$ in $f_{\mu}$ and $B_{\mu}$
in (\ref{dualCFJgen}) for example, by the duality procedure ($D_{\mu}$ is purely
topological and carries the symmetry aspects of $A_\mu$) may cause a certain uneasiness and questions arise as if it is possible
to write the dynamical content of the dual lagrangian with just one field.
That seems to be expected since their duals were obtained from a simple vectorial potential just by singling out the direction determined by the Lorentz-breaking vector.
This is
not a trivial matter and needs some elaboration. Looking at our starting point
(\ref{mestraCFJ}) we might be tempted to naively identify $\Pi_{\mu\nu}$ with the
dual field of $A_{\mu}$.
Indeed, the discussion at the begining of this section seems to indicate that those fields should in fact be the components of a rank-two potential, as the massive duality relation, $p+q+1=D$, suggests for this 4-dimensional example. Therefore a sort of constrained self-dual rank-two model should be the result of this duality transformation.
Of course  this can only be done if there are constraints which reduce the independent components of $\Pi_{\mu\nu}$ (so that we
can preserve the original two degrees of freedom described by $A_{\mu}$).

In fact a direct dual of the CFJ-model can also be obtained with another method which involves the
elimination in the first-order lagrangian (Eq.(\ref{mestraCFJ}) in this case) of the
original dynamical field $A^{\mu}$ in favor of the auxiliary field
$\Pi^{\mu\nu}$. The action when written in terms of just this
auxiliary field seems to be the dual theory we are looking for. Let us see how this works for the
Carroll-Field-Jackiw model and then show that the result indeed decomposes as in (\ref{dualCFJgen}), as it should.
Consider again the first-order lagrangian (\ref{mestraCFJ}).
%\begin{eqnarray}
% {\cal L}_{MCFJ} = \frac 12 \Pi_{\mu\nu}\varepsilon^{\mu\nu\rho\sigma}\partial_{\rho}A_{\sigma} - \frac 14 \Pi_{\mu\nu}\Pi^{\mu\nu} +
% p_\mu\varepsilon^{\mu\nu\rho\sigma}A_{\nu}\partial_{\rho}A_{\sigma}.
%    \label{mestraCFJapB}
%\end{eqnarray}
Since this is a massive theory, albeit gauge invariant, the auxiliary field will end up playing the role of the dual rank-two potential.
The Euler-Langrange equations for the $A^{\mu}$ field gives
\begin{eqnarray}
 \Lambda^{\mu} \equiv \varepsilon^{\mu\nu\rho\sigma}\partial_{\nu}\Pi_{\rho\sigma} = 4 \varepsilon^{\mu\nu\rho\sigma}p_\nu\partial_{\rho}A_{\sigma}.
    \label{lambda}
\end{eqnarray}
Observe that $\Lambda^{\mu}$ satisfies
\begin{eqnarray}
 \partial_{\mu}\Lambda^{\mu} = 0
    \label{lambda1}
\end{eqnarray}
by definition and
\begin{eqnarray}
 p_{\mu}\Lambda^{\mu} = 0
    \label{lambda2}
\end{eqnarray}
in virtue of the equation of motion obeyed by $A^{\mu}$. Formally, substituting
(\ref{lambda}) in (\ref{mestraCFJ}) we have
\begin{eqnarray}
 {\cal L}_{MCFJ} \rightarrow \frac 14 \Lambda^{\mu}A_{\mu} - \frac 14 \Pi_{\mu\nu}\Pi^{\mu\nu}.
    \label{mestraCFJapB2}
\end{eqnarray}
where $A^{\mu}\equiv A^{\mu}(\Pi)$ is defined by (\ref{lambda}). To complete the procedure we have to
rewrite $\Lambda^{\mu}A_{\mu}$ as a function of $\Pi^{\mu\nu}$. It follows from
(\ref{lambda}) that the solution of the constraint
\begin{eqnarray}
 \varepsilon^{\mu\nu\rho\sigma}\partial_{\nu}(\Pi_{\rho\sigma} + 2 p_{[\rho}A_{\sigma]}) =
 0\nonumber\\
 \Rightarrow \Pi_{\mu\nu} = - 2 p_{[\mu}A_{\nu]} + \partial_{[\mu}B_{\nu]},
    \label{defPi}
\end{eqnarray}
introduces a new vector potential $B^{\mu}$. Hence
\begin{eqnarray}
 p^{\mu}\Lambda^{\nu}\Pi_{\mu\nu} = - 2 p^2\Lambda^{\mu}A_{\mu} + \Lambda^{\mu}(p^{\nu}\partial_{\nu})B_{\mu} - \Lambda^{\mu}\partial_{\mu}(p^{\nu}B_{\nu}),
    \label{Pirel}
\end{eqnarray}
where (\ref{lambda2}) was used. As we substitute this in (\ref{mestraCFJapB2}) we can
integrate by parts and the last term of (\ref{Pirel}) drops out because of
(\ref{lambda1}) and we are left with
\begin{eqnarray}
 {\cal L}_{dualCFJ} &=& -\frac{1}{8p^2}(p^{\alpha}\Pi_{\alpha\mu})\varepsilon^{\mu\nu\rho\sigma}\partial_{\nu}\Pi_{\rho\sigma} - \frac 14
 \Pi_{\mu\nu}\Pi^{\mu\nu} + \frac{1}{8p^2} [(p^{\alpha}\partial_{\alpha})B_{\mu}]\varepsilon^{\mu\nu\rho\sigma}\partial_{\nu}\Pi_{\rho\sigma}.
    \label{dualCFJgenapb}
\end{eqnarray}
Observe that the relation (\ref{defPi}) does not contain the component of $A_{\mu}$
in the direction defined by $p^{\mu}$. In fact this component is a Lagrange
multiplier in (\ref{mestraCFJ}). However, the constraint imposed by this component
establishes a relation between $\Pi_{\mu\nu}$ and $B_{\mu}$ that can be read from
(\ref{defPi}) as
\begin{eqnarray}
 \varepsilon^{\mu\nu\rho\sigma}p_{\nu}\Pi_{\rho\sigma} =
 \varepsilon^{\mu\nu\rho\sigma}p_{\nu}\partial_{\rho}B_{\sigma}.
    \label{Pirel2}
\end{eqnarray}
This is the origin of the split of the fields. $\Pi_{\mu\nu}$ is indeed the dual of
$A_{\mu}$ but its components must satisfy certain constraints that are more easily
dealt with if we introduce the $B_{\mu}$ field. Proceeding further, observe that\begin{eqnarray}
(\varepsilon^{\mu\nu\rho\sigma}p_{\nu}\Pi_{\rho\sigma})^2 = 2p^2
\Pi_{\mu\nu}\Pi^{\mu\nu} - 4
(p^{\alpha}\Pi_{\alpha\mu})(p_{\beta}\Pi^{\beta\mu})\nonumber\\
\Rightarrow \Pi_{\mu\nu}\Pi^{\mu\nu} = \frac{2}{p^2}
(\varepsilon^{\mu\nu\rho\sigma}p_{\nu}\partial_{\rho}B_{\sigma})^2 + \frac{2}{p^2}
(p^{\alpha}\Pi_{\alpha\mu})(p_{\beta}\Pi^{\beta\mu}),
   \label{Pirel3}
\end{eqnarray}
where we have used (\ref{Pirel2}). As for the first term of (\ref{dualCFJgenapb}), it
can be rearranged as
\begin{eqnarray}
-\frac{1}{8p^2}(p^{\alpha}\Pi_{\alpha\mu})\varepsilon^{\mu\nu\rho\sigma}\partial_{\nu}\Pi_{\rho\sigma}
= -\frac{1}{4p^4}
(p^{\alpha}\Pi_{\alpha\mu})\varepsilon^{\mu\nu\rho\sigma}p_{\nu}\partial_{\rho}[(p^{\beta}\partial_{\beta})B_{\sigma}]
+ \frac{1}{4p^4}
(p^{\alpha}\Pi_{\alpha\mu})\varepsilon^{\mu\nu\rho\sigma}p_{\nu}\partial_{\rho}(p^{\beta}\Pi_{\beta\sigma}),
   \label{Pirel4}
\end{eqnarray}
and similarly for the last term
\begin{eqnarray}
\frac{1}{8p^2}
[(p^{\alpha}\partial_{\alpha})B_{\mu}]\varepsilon^{\mu\nu\rho\sigma}\partial_{\nu}\Pi_{\rho\sigma}
= \frac{1}{4p^4}
[(p^{\alpha}\partial_{\alpha})B_{\mu}]\varepsilon^{\mu\nu\rho\sigma}p_{\nu}\partial_{\rho}[(p^{\beta}\partial_{\beta})B_{\sigma}]
- \frac{1}{4p^4}
[(p^{\alpha}\partial_{\alpha})B_{\mu}]\varepsilon^{\mu\nu\rho\sigma}p_{\nu}\partial_{\rho}(p^{\beta}\Pi_{\beta\sigma})\, .
   \label{Pirel5}
\end{eqnarray}
Gathering these results, (\ref{dualCFJgenapb}) can be written as
\begin{eqnarray}
 {\cal L}_{dualCFJ} &=&  \frac{1}{4p^4} (p^{\alpha}\Pi_{\alpha\mu})\varepsilon^{\mu\nu\rho\sigma}p_{\nu}\partial_{\rho}(p^{\beta}\Pi_{\beta\sigma}) - \frac{1}{2p^2} (p^{\alpha}\Pi_{\alpha\mu}) (p_{\beta}\Pi^{\beta\mu}) - \frac{1}{2p^2} (\varepsilon^{\mu\nu\rho\sigma}p_{\nu}\partial_{\rho}B_{\sigma})^2\nonumber\\
  &+& \frac{1}{2p^4}
  (p^{\alpha}\Pi_{\alpha\mu})\varepsilon^{\mu\nu\rho\sigma}p_{\nu}\partial_{\rho}[(p^{\beta}\partial_{\beta})B_{\sigma}]
 + \frac{1}{4p^4}[(p^{\alpha}\partial_{\alpha})B_{\mu}]\varepsilon^{\mu\nu\rho\sigma}p_{\nu}\partial_{\rho}[(p^{\beta}\partial_{\beta})B_{\sigma}].
    \label{dualCFJgenapb2}
\end{eqnarray}
which is just (\ref{dualCFJgen}) if we define $p^{\alpha}\Pi_{\alpha\mu} \equiv
f_{\mu}$.

\section{Dual with sources}

In this section we will generalize the results of the last section by considering the
CFJ model minimally coupled with a conserved external source
\begin{eqnarray}
\label{CFJJ}
  {\cal L}^{J}_{CFJ} &=& - \frac 14 F_{\mu\nu}F^{\mu\nu} +
  p_{\mu}\varepsilon^{\mu\nu\rho\sigma}A_{\nu}\partial_{\rho}A_{\sigma} - eJ_{\mu}A^{\mu}\, ,
\end{eqnarray}
which may be rewritten with the auxiliary field $\Pi^{\mu\nu}$ that, as usual, will play the role of dual field, leading to
\begin{eqnarray}
 {\cal L}^J_{MCFJ} = \frac 12 \Pi_{\mu\nu}\varepsilon^{\mu\nu\rho\sigma}\partial_{\rho}A_{\sigma} - \frac 14 \Pi_{\mu\nu}\Pi^{\mu\nu} +
 p_\mu\varepsilon^{\mu\nu\rho\sigma}A_{\nu}\partial_{\rho}A_{\sigma} - eJ_{\mu}A^{\mu}.
    \label{mestraCFJJ}
\end{eqnarray}
As before, for simplicity, we will work with $p^{\mu} = (0,0,0,m)$. Further, as
$J^{\mu}$ is conserved, we may write it using a so called Chern-kernel
$\Lambda_{\mu\nu}$ so that (see appendix A for further discussion of this concept)
\begin{eqnarray}
\label{ck}
  J^{\mu} = \frac 12
  \varepsilon^{\mu\nu\rho\sigma}\partial_{\nu}\Lambda_{\rho\sigma}
\end{eqnarray}
or in terms of its components
\begin{eqnarray}
\label{ck2}
    J^a &=& \frac 12  \varepsilon^{abc}\partial_{3}\Lambda_{bc}-\varepsilon^{abc}\partial_{b}\omega_{c}\\
    J^3 &=& -\frac 12 \varepsilon^{abc}\partial_{a}\Lambda_{bc}.
\end{eqnarray}
From its definition we see that $\Lambda_{\mu\nu}$ possess some freedom,
that is, $J^{\mu}$ is invariant under
\begin{eqnarray}
 \label{ck3}
    \Lambda_{\mu\nu} \rightarrow \Lambda_{\mu\nu} + \partial_{\mu}H_{\nu} - \partial_{\nu}H_{\mu} .
\end{eqnarray}
The Lagrangian (\ref{mestraCFJJ}) becomes
\begin{eqnarray}
  {\cal L}^{J}_{MCFJ} &=& \frac 12 A_3 (-\varepsilon^{abc}\partial_{a}\Pi_{bc} - 2eJ^3) + \frac 12 A_{a}\varepsilon^{a\nu\rho\sigma}\partial_{\rho}\Pi_{\rho\sigma}
  - \frac 14 \Pi_{\mu\nu}\Pi^{\mu\nu} + m\varepsilon^{abc}A_{a}\partial_{b}A_{c} - eJ_{a}A^{a}.
\end{eqnarray}
Again $A^3$ is a Lagrange multiplier which enforces a constraint
\begin{eqnarray}
 -\varepsilon^{abc}\partial_{a}\Pi_{bc} + e\varepsilon^{abc}\partial_{a}\Lambda_{bc} = 0
    \label{lagmultJ}
\end{eqnarray}
which changes the solution of (\ref{lagmult2}) to include the presence of the source
\begin{eqnarray}
\Pi_{ab}= \partial_{a}B_{b} -
 \partial_{b}B_{a} + e\Lambda_{ab} \equiv \bar{G}_{ab}
    \label{lagmultJ2}
\end{eqnarray}
displaying now a non-minimal (or magnetic) coupling. This coupling, as is well known, modifies the Bianchi identity to include $J_{\mu}$ as a magnetic current. Then we have
\begin{eqnarray}
\label{CFJJm}
 {\cal L}^{J}_{MCFJ} &=& \frac 12 A_{a}\varepsilon^{abc}\partial_{3}\bar{G}_{bc} - A_{a}\varepsilon^{abc}\partial_{b}h_{c}- \frac 14 \bar{G}_{ab}\bar{G}^{ab}
 + \frac 12 h_ah^a + m\varepsilon^{abc}A_{a}\partial_{b}A_{c}- eJ_{a}A^{a}\nonumber\\
               &=& (-h_a + e\omega _a + \partial_{3}B_a + mA_a)\varepsilon^{abc}\partial_{b}A_{c}- \frac 14 \bar{G}_{ab}\bar{G}^{ab}
 + \frac 12 h_ah^a.
\end{eqnarray}
where again we have renamed the independent field $\Pi^{a3} \equiv h^a$ and used (\ref{ck}). From here everything follows as before, after (\ref{mestraCFJ3}), but now with $h^a \rightarrow h^a -e\omega^a$.
Nevertheless it is interesting to see how the presence of the Chern-kernel affects
the procedure. Consider next the field redefinition
\begin{eqnarray}
-(h_a - e\omega_a) + mA_a = C_a
    \label{redefJ}
\end{eqnarray}
leading to
\begin{eqnarray}
 {\cal L}^J_{MCFJ} \rightarrow C_a\varepsilon^{abc}\partial_{b}A_{c} + \partial_{3}B_a\varepsilon^{abc}\partial_{b}A_{c} - \frac 14 \bar{G}_{ab}\bar{G}^{ab}
 + \frac 12 (mA_a + e\omega_a - C_a)(mA^a + e\omega^a - C^a).
    \label{mestraCFJJ4}
\end{eqnarray}
Observe that now $C_a$ must inherit the gauge symmetry not only of $A_a$ but of $\omega_a$ as well
as so that the last term above is symmetric under $A_a \rightarrow A_a
+ \partial_a\phi$ and $\omega_a \rightarrow \omega_a + \partial_a\psi$ with $C_a
\rightarrow C_a + m\partial_a\phi +e\partial_a\psi$. Now we perform a canonical transformation in
field space
\begin{eqnarray}
C_{a} &=& \frac 12 (A^{+}_{a} - A^{-}_{a})\nonumber\\
   A_{a} &=& \frac{1}{2m} (A^{+}_{a} + A^{-}_{a}),
    \label{rotJ}
\end{eqnarray}
which gives us
\begin{eqnarray}
 {\cal L}' &=& \frac{1}{4m} A^{+}_a\varepsilon^{abc}\partial_{b}A^{+}_{c} - \frac{1}{4m} A^{-}_a\varepsilon^{abc}\partial_{b}A^{-}_{c} + \frac{1}{2m}\partial_{3}B_a\varepsilon^{abc}\partial_{b}A^{+}_{c} + \frac{1}{2m}\partial_{3}B_a\varepsilon^{abc}\partial_{b}A^{-}_{c}\nonumber\\
  &-& \frac 14 \bar{G}_{ab}\bar{G}^{ab}
 + \frac 12 (A^{-}_a+e\omega_a) (A^{-a}+e\omega^a).
    \label{mestraCFJJ5}
\end{eqnarray}
We can make a further field redefinition to separate out the pure-CS term
\begin{eqnarray}
A^{+}_{a} + \partial_{3}B_a = D_a \, ,
    \label{redefJ2}
\end{eqnarray}
which leads us to
\begin{eqnarray}
 {\cal L}_{dualCFJ} &=& \frac{1}{4m} D_a\varepsilon^{abc}\partial_{b}D_{c} - \frac{1}{4m} A^{-}_a\varepsilon^{abc}\partial_{b}A^{-}_c + \frac 12 (A^{-}_a - e\omega_{a}) (A^{-a} - e\omega^{a}) - \frac 14 \bar{G}_{ab}\bar{G}^{ab} + \frac{1}{2m} A^{-}_a\varepsilon^{abc}\partial_{b}\partial_{3}B_{c}\nonumber\\
  &-& \frac{1}{4m}\partial_{3}B_a\varepsilon^{abc}\partial_{b}\partial_{3}B_{c}.
    \label{dualCFJJ}
\end{eqnarray}
Observe that as it stands this Lagrangian tells us that $A^{-}_a$ is a gauge field
with the symmetry necessary to maintain the freedom provided by $\omega_a$. With the
Stuckelberg-like redefinition
\begin{eqnarray}
A^{-}_a - e\omega_{a} = f_a
    \label{redefJ3}
\end{eqnarray}
we may write (\ref{dualCFJJ}) in terms of the ``self-dual-like'', non-gauge field
$f_a$
\begin{eqnarray}
 {\cal L}^J_{dualCFJ} = \frac{1}{4m} D_a\varepsilon^{abc}\partial_{b}D_{c} - \frac{1}{4m} (f_a - e\omega_{a} - \partial_{3}B_a)\varepsilon^{abc}\partial_{b}(f_c - e\omega_{c} - \partial_{3}B_c) + \frac 12 f_a f^a - \frac 14 \bar{G}_{ab}\bar{G}^{ab}.
    \label{dualCFJJM}
\end{eqnarray}
%with $\bar{G}_{ab} = \partial_{a}B_{b} - \partial_{b}B_{a} + e\Lambda_{ab}$.
As expected, both the Maxwell and the self-dual fields induced by the duality transformation display a non-minimal coupling with the original source.

\vspace{0.4in}
\section{Conclusion}

We have obtained the dual formulation of the Carroll-Field-Jackiw model as dictated
by distinct duality techniques. In this respect both the dual projection approach as well as the standard master action duality technique, obtained by solving the equations of motion of the Legendre transformed action, were used, giving complementary information. The asymmetry introduced with the $p_{\mu}$ vector is
pronounced by the duality procedure inducing a ``splitting'' in components of the
fields. This splitting of the CFJ-model in terms of a Maxwell-Chern-Simons and a scalar field is better dealt with in the dual projection approach, including the couplings to external charges. The dualization procedure induces a splitted action in terms of a self-dual and a Maxwell field, respectivelly.
We have also establish the duality in the presence of a external source. With the
help of the concept of the Chern-kernel we were able to see the splitting of the
coupling terms as a consequence of the Lorentz breaking.
In this interacting situation, the original action shows an electrical coupling while duality induces a magnetic coupling, as expected.

We have seen that part of the gauge symmetry survives the duality procedure (as
represented by the $B_a$ field in (\ref{dualCFJ}), for example) but the other part
decouples (represented in (\ref{dualCFJ}) by $D_a$). Therefore, similarly to the $(2+1)D$
MCS-SD duality, we are led to the conclusion that, due to the existence of a pure-CS term, a topological degeneracy is present in the dual of the Carroll-Field-Jackiw model as well.

It was also found, through another technique, that a direct dual of the CFJ-model is given in terms of a single rank-two potential, albeit constrained. As discussed in the main text this was to be expected from general arguments of massive versus massless duality in diverse dimensions. Since this CFJ-model is to be seem as the photonic sector of an extended standard model, it would be interesting to consider the possibility of couplings these modes with dynamical fermionic and bosonic matter. It would also be quite interesting to consider the duality of this model when the CFJ potential $A_\mu$ is coupled to a magnetic monopole or, what is naturally expected, when charges both of electric and magnetic nature are present simultaneously. Research in these directions are presently in progress.

\section{Appendix A}

In this appendix we will recall some results originally obtained in \cite{nosso}
concerning the Maxwell-Chern-Simons/Self-Dual duality when the MCS theory is
minimally coupled with an external current. Before we engage in the actual
dualization procedure we would like to call attention to the Chern-kernel concept
which is quite useful in the presence of magnetic charges or, as it is the case here,
in the dualization procedure that interchanges the notion of magnetic and electric
charges. Whenever we are given a conserved current $J^{\mu}$ such that
\begin{eqnarray}
 \partial_{\mu}J^{\mu} = 0
    \label{appA01}
\end{eqnarray}
a Chern-kernel $\Lambda_{\rho\sigma}$ may naturally be defined by the formal solution of (\ref{appA01})
\begin{eqnarray}
  J^{\mu} = \frac 12 \varepsilon^{\mu\nu\rho\sigma}
  \partial_{\nu}\Lambda_{\rho\sigma}\, .
    \label{appA02}
\end{eqnarray}
For the sake of this argument we work in $(3+1)D$, but the generalization to any number
of dimensions is straightforward (see below).
It is important to note that
$\Lambda_{\mu\nu}$ is non-observable. In fact this object is the natural generalization of the notion of Dirac string for extended charges. We see clearly from (\ref{appA02}) that there is a ambiguity in its definition. The
transformation
\begin{eqnarray}
  \Lambda_{\mu\nu} \rightarrow \Lambda_{\mu\nu} + \partial_{\mu}C_{\nu} - \partial_{\nu}C_{\mu}
    \label{appA03}
\end{eqnarray}
indeed leaves $J^{\mu}$ unchanged. If $J^{\mu}$ is minimally coupled to an abelian gauge field
$A_{\mu}$, its conservation (\ref{appA01}) is a necessary condition for the
maintenance of the gauge symmetry of the interaction term. In this sense (\ref{appA03}) is directly
related to the gauge transformation of $A_{\mu}$.

The relevance of the Chern-kernel concept reveals itself when we dualize a gauge
theory minimally coupled with $J^{\mu}$. Take for instance the Maxwell theory,
\begin{eqnarray}
  {\cal L}_{Maxwell}= -\frac 14 F_{\mu\nu}F^{\mu\nu} - eA_{\mu}J^{\mu}.
  \label{appA04}
\end{eqnarray}
Upon duality the dynamics becomes described by the dual potential $\tilde{A}_{\mu}$
while the coupling becomes non-minimal
\begin{eqnarray}
  {\cal L}_{Maxwell} \rightarrow -\frac 14 \tilde{F}_{\mu\nu}\tilde{F}^{\mu\nu}
  \label{appA05}
\end{eqnarray}
with
\begin{eqnarray}
  \tilde{F}_{\mu\nu} = \partial_{\mu}\tilde{A}_{\nu} - \partial_{\nu}\tilde{A}_{\mu}
  - e\, \Lambda_{\mu\nu}.
  \label{appA06}
\end{eqnarray}
This amounts to a modification of the Bianchi identity making $J^{\mu}$ a magnetic
current in this dual formulation
\begin{eqnarray}
  \partial_{\mu}\tilde{F}^{\mu\nu} = e \, J^{\nu}
  \label{appA07}
\end{eqnarray}
where $\tilde{F}^{\mu\nu} = \frac 12 \varepsilon^{\mu\nu\rho\sigma}
  \partial_{\nu}\tilde{F}_{\rho\sigma}$. This is very well known of course \cite{dirac}
but the point here is to stress the generality of the Chern-kernel concept and its
fundamental importance in the dual formulation of interacting gauge theories.

We are now ready to consider the MCS/SD duality including couplings \cite{nosso} which will teach us how the self-dual model couples magnetically (non-minimally) with the sources.
Consider the MCS theory minimally coupled with a conserved current
\begin{eqnarray}
  {\cal L}^{J}_{MCS}= -\frac 14 F_{\mu\nu}F^{\mu\nu} + \frac m2 A_{\mu}\varepsilon^{\mu\nu\rho}\partial_{\nu}A_{\rho} - eA_{\mu}J^{\mu},
  \label{appA08}
\end{eqnarray}
which may be written as
\begin{eqnarray}
 {\cal L}^{J}_{MCS} &\rightarrow& \Pi_{\mu}(\varepsilon^{\mu\nu\rho}\partial_{\nu}A_{\rho}) + \frac 12 \Pi_{\mu}\Pi^{\mu} + \frac m2 A_{\mu}\varepsilon^{\mu\nu\rho}\partial_{\nu}A_{\rho} - eA_{\mu}\varepsilon^{\mu\nu\rho}\partial_{\nu}\omega_{\rho}\nonumber\\
              &=&(\frac m2 A_{\mu} + \Pi_{\mu} -e\omega_{\mu})(\varepsilon^{\mu\nu\rho}\partial_{\nu}A_{\rho}) + \frac 12 \Pi_{\mu}\Pi^{\mu},
    \label{appA09}
\end{eqnarray}
where $\Pi_\mu$ is an auxiliary field which may be integrated out to give us back the MCS
lagrangian and we have used the appropriate definition of the Chern-kernel for
$(2+1)D$, that is
\begin{eqnarray}
 J^{\mu} = \varepsilon^{\mu\nu\rho}\partial_{\nu}\omega_{\rho}
   \label{appA10}
\end{eqnarray}
By making the redefinition $\frac m2 A_{\mu} + \Pi_{\mu} - e\omega_{\mu} = B_{\mu}$,
we find for (\ref{appA09})
\begin{eqnarray}
{\cal L} = B_{\mu}(\varepsilon^{\mu\nu\rho}\partial_{\nu}A_{\rho}) + \frac 12
(B_{\mu} - \frac m2 A_{\mu} -e\omega_{\mu})(B^{\mu} - \frac m2 A^{\mu}
-e\omega^{\mu}).
    \label{appA11}
\end{eqnarray}
Everything follows very similarly with the free case, but it is interesting
to note the role of the new symmetry introduced by the Chern-kernel. As before
observe that by definition $B_{\mu}$ transforms as $B_{\mu} \rightarrow B_{\mu} +
\frac m2
\partial_{\mu}\phi$ whenever $A_{\mu} \rightarrow A_{\mu} +
\partial_{\mu}\phi$, but it must also inherit the symmetry of $\omega_{\mu}$, that
is, $B_{\mu}$ transforms as $B_{\mu} \rightarrow B_{\mu} - e
\partial_{\mu}\psi$ whenever $\omega_{\mu} \rightarrow \omega_{\mu} +
\partial_{\mu}\psi$.

Next we perform the usual canonical transformation in the space of the fields
\begin{eqnarray}
 B_{\mu} &=& \frac 12 (A^{+}_{\mu} - A^{-}_{\mu})\nonumber\\
   A_{\mu} &=& \frac 1m (A^{+}_{\mu} + A^{-}_{\mu}),
    \label{appA12}
\end{eqnarray}
The transformations discussed above are encoded in the new fields $A^{+}_{\mu}$ and
$A^{-}_{\mu}$ as follows.

\noindent As $A_{\mu} \rightarrow A_{\mu} +
\partial_{\mu}\phi$,
\begin{eqnarray}
 A^{+}_{\mu} &\rightarrow& A^{+}_{\mu} + \frac m2 \partial_{\mu}\phi\nonumber\\
   A^{-}_{\mu} &\rightarrow& A^{-}_{\mu},
    \label{appA13}
\end{eqnarray}
and as $\omega_{\mu} \rightarrow \omega_{\mu} +
\partial_{\mu}\psi$,
\begin{eqnarray}
 A^{+}_{\mu} &\rightarrow& A^{+}_{\mu} - e
\partial_{\mu}\psi\nonumber\\
   A^{-}_{\mu} &\rightarrow& A^{-}_{\mu} + e
\partial_{\mu}\psi.
    \label{appA14}
\end{eqnarray}

Rewriting the Lagrangian (\ref{appA11}) using (\ref{appA12}) we obtain
\begin{eqnarray}
{\cal L}_{dual} = \frac{1}{2m}
A^{+}_{\mu}\varepsilon^{\mu\nu\rho}\partial_{\nu}A^{+}_{\rho} - \frac{1}{2m}
A^{-}_{\mu}\varepsilon^{\mu\nu\rho}\partial_{\nu}A^{-}_{\rho} + \frac 12 (A^{-}_{\mu}
- e\omega_{\mu})(A^{-\mu} - e\omega^{\mu}) ,
    \label{appA15}
\end{eqnarray}
or, with the redefinition $A^{-}_{\mu} - e\omega_{\mu} \equiv f_{\mu}$
\begin{eqnarray}
{\cal L}_{dual} = \frac{1}{2m}
A^{+}_{\mu}\varepsilon^{\mu\nu\rho}\partial_{\nu}A^{+}_{\rho} - \frac{1}{2m} (f_{\mu}
+ e\omega_{\mu})\varepsilon^{\mu\nu\rho}\partial_{\nu}(f_{\rho} + e\omega_{\rho}) +
\frac 12 f_{\mu}f^{\mu}.
    \label{appA16}
\end{eqnarray}
Observe that $f^{\mu}$ is invariant under both transformations described in
(\ref{appA13}) and (\ref{appA14}). This result shows that the pure Chern-Simons term
remain decoupled from the sources and reveals the way in which the self-dual field
$f^{\mu}$ experiences the e-charge as a ``magnetic'' one (interpreting the original
coupling with $A_{\mu}$ in the MCS theory as electric).

\section{Appendix B}

Here we will consider in detail the derivation of the dual of the CFJ model for
general non-light-like $p^{\mu}$. We start with the first-order lagrangian
\begin{eqnarray}
 {\cal L}_{MCFJ} = \frac 12 \Pi_{\mu\nu}\varepsilon^{\mu\nu\rho\sigma}\partial_{\rho}A_{\sigma} - \frac 14 \Pi_{\mu\nu}\Pi^{\mu\nu} +
 p_\mu\varepsilon^{\mu\nu\rho\sigma}A_{\nu}\partial_{\rho}A_{\sigma}.
    \label{mestraCFJap}
\end{eqnarray}
We can single out the
components which are in the $p^{\mu}$ direction, obtaining
\begin{eqnarray}
 {\cal L}_{MCFJ} &=& \frac{1}{p^2} p_{\mu} (p^{\alpha}\Pi_{\alpha\nu})\varepsilon^{\mu\nu\rho\sigma}\partial_{\rho}A_{\sigma} +  \frac{1}{2p^2} \Pi_{\mu\nu}\varepsilon^{\mu\nu\rho\sigma}p_{\rho}(p^{\alpha}\partial_{\alpha})A_{\sigma} +  \frac{1}{2p^2} \Pi_{\mu\nu}\varepsilon^{\mu\nu\rho\sigma}\partial_{\rho}p_{\sigma}(p^{\alpha}A_{\alpha})\nonumber\\
  &-& \frac{1}{8p^2} (\varepsilon^{\mu\nu\rho\sigma}p_{\nu}\Pi_{\rho\sigma})^2 -\frac{1}{2p^2}(p^{\alpha}\Pi_{\alpha\mu})^2 +
 p_\mu\varepsilon^{\mu\nu\rho\sigma}A_{\nu}\partial_{\rho}A_{\sigma}\, ,
    \label{mestraCFJap2}
\end{eqnarray}
where we used the fact that
\begin{eqnarray}
 \Pi_{\mu\nu}\Pi^{\mu\nu} = \frac{1}{2p^2} (\varepsilon^{\mu\nu\rho\sigma}p_{\nu}\Pi_{\rho\sigma})^2 + \frac{2}{p^2}(p^{\alpha}\Pi_{\alpha\mu})^2.
    \label{piquad}
\end{eqnarray}
We see that $p^{\alpha}A_{\alpha}$ is a lagrange multiplier which enforces
\begin{eqnarray}
 \varepsilon^{\mu\nu\rho\sigma}p_{\mu}\partial_{\nu}\Pi_{\rho\sigma} = 0 \Rightarrow \varepsilon^{\mu\nu\rho\sigma}p_{\nu}\Pi_{\rho\sigma} = 2\varepsilon^{\mu\nu\rho\sigma}p_{\nu}\partial_{\rho}B_{\sigma}
 \label{lagmult55}
\end{eqnarray}
and we are left with
\begin{eqnarray}
 {\cal L}_{MCFJ} &=& \frac{1}{p^2} [ -(p^{\alpha}\Pi_{\alpha\mu}) - p^2 A_{\mu} + (p^{\alpha}\partial_{\alpha})B_{\mu}]\varepsilon^{\mu\nu\rho\sigma}p_{\nu}\partial_{\rho}A_{\sigma}\nonumber\\
  &-& \frac{1}{2p^2} (\varepsilon^{\mu\nu\rho\sigma}p_{\nu}\partial_{\rho}B_{\sigma})^2 -\frac{1}{2p^2}(p^{\alpha}\Pi_{\alpha\mu})^2.
    \label{mestraCFJap3}
\end{eqnarray}
We can make a field redefinition
\begin{eqnarray}
-(p^{\alpha}\Pi_{\alpha\mu}) - p^2 A_{\mu} = C_{\mu}
    \label{fieldred}
\end{eqnarray}
Observe that even though we have used the $\mu$ indices, $C_{\mu}$ does not have a
component in the $p^{\mu}$ direction (as we already eliminated $p^{\alpha}A_{\alpha}$
and $\Pi_{\mu\nu}$ is antisymmetric) this property will remain true for all fields
redefinitions that will follow. With this redefinition we get
\begin{eqnarray}
 {\cal L}_{MCFJ} &\rightarrow& \frac{1}{p^2}C_{\mu}\varepsilon^{\mu\nu\rho\sigma}p_{\nu}\partial_{\rho}A_{\sigma} + \frac{1}{p^2}(p^{\alpha}\partial_{\alpha})B_{\mu}\varepsilon^{\mu\nu\rho\sigma}p_{\nu}\partial_{\rho}A_{\sigma}\nonumber\\
  &-& \frac{1}{2p^2} (\varepsilon^{\mu\nu\rho\sigma}p_{\nu}\partial_{\rho}B_{\sigma})^2 -\frac{1}{2p^2}(C_{\mu} + p^2 A_{\mu})^2.
    \label{mestraCFJap4}
\end{eqnarray}
We make a further redefinition, as usual,
\begin{eqnarray}
C_{\mu} &=& \frac 12 (A^{+}_{\mu} - A^{-}_{\mu})\nonumber\\
   A_{\mu} &=& \frac{1}{2p^2} (A^{+}_{\mu} + A^{-}_{\mu}),
    \label{fieldred2}
\end{eqnarray}
which gives us
\begin{eqnarray}
 {\cal L}_{MCFJ} &\rightarrow& \frac{1}{4p^4}A^{+}_{\mu}\varepsilon^{\mu\nu\rho\sigma}p_{\nu}\partial_{\rho}A^{+}_{\sigma} - \frac{1}{4p^4}A^{-}_{\mu}\varepsilon^{\mu\nu\rho\sigma}p_{\nu}\partial_{\rho}A^{-}_{\sigma} + \frac{1}{2p^4}(p^{\alpha}\partial_{\alpha})B_{\mu}\varepsilon^{\mu\nu\rho\sigma}p_{\nu}\partial_{\rho}A^{+}_{\sigma}\nonumber\\
  &+& \frac{1}{2p^4}(p^{\alpha}\partial_{\alpha})B_{\mu}\varepsilon^{\mu\nu\rho\sigma}p_{\nu}\partial_{\rho}A^{-}_{\sigma}
  - \frac{1}{2p^2} (\varepsilon^{\mu\nu\rho\sigma}p_{\nu}\partial_{\rho}B_{\sigma})^2 -\frac{1}{2p^2}(A^{+}_{\mu})^2.
    \label{mestraCFJap5}
\end{eqnarray}
Defining next
\begin{eqnarray}
D_{\mu} &=& (p^{\alpha}\partial_{\alpha})B_{\mu} - A^{-}_{\mu}
    \label{fieldred3}
\end{eqnarray}
we finally get, with $A^{+}_{\mu}=f_{\mu}$,
\begin{eqnarray}
 {\cal L}_{dualCFJ} &=& -\frac{1}{4p^4} D_{\mu}\varepsilon^{\mu\nu\rho\sigma}p_{\nu}\partial_{\rho}D_{\sigma} + \frac{1}{4p^4} f_{\mu}\varepsilon^{\mu\nu\rho\sigma}p_{\nu}\partial_{\rho}f_{\sigma} - \frac{1}{2p^2} f_{\mu} f^{\mu} - \frac{1}{2p^2} (\varepsilon^{\mu\nu\rho\sigma}p_{\nu}\partial_{\rho}B_{\sigma})^2\nonumber\\
  &+& \frac{1}{2p^4}
  f_{\mu}\varepsilon^{\mu\nu\rho\sigma}p_{\nu}\partial_{\rho}[(p^{\alpha}\partial_{\alpha})B_{\sigma}]
 + \frac{1}{4p^4}[(p^{\alpha}\partial_{\alpha})B_{\mu}]\varepsilon^{\mu\nu\rho\sigma}p_{\nu}\partial_{\rho}[(p^{\beta}\partial_{\beta})B_{\sigma}].
    \label{dualCFJgenapp}
\end{eqnarray}
which is the desired result, showing the pure-CS field $D_\mu$, the self-dual field $f_\mu$ and the Maxwell field, $B_\mu$, besides their interaction pieces.

\vspace{0.3in} \noindent {\bf Acknowledgments} This work is partially supported by
PROCAD/CAPES and PRONEX/CNPq. MSG and LG thanks CNPq and CAPES respectively
(Brazilian research agencies) for financial support.

\end{document}